\documentclass[12pt,aps,prd,preprint,groupedaddress,tightenlines,nofootinbib,floatfix]{revtex4}
\usepackage[dvipdfm]{graphicx}

\newcommand{\reseteqnum}{\setcounter{equation}{0}}
\newcommand{\nn}{\nonumber}

\newcommand{\ovl}[1]{\overline{#1}}
\newcommand{\wt}[1]{\widetilde{#1}}

\newcommand{\eqn}[1]{(\ref{#1})}
\newcommand{\p}{\partial}
\newcommand{\bpsi}{{\overline{\psi}}}

\newcommand{\pslash}{p\kern-1ex /}
\newcommand{\Dslash}{{\cal D}\kern-1.5ex /}
\newcommand{\tr}{{\rm tr}}

\newcommand{\msbar}{\overline{\rm MS}}
\begin{document}

\begin{flushright}
{\normalsize UTHEP-584}\\
{\normalsize UTCCS-P-54}\\
\end{flushright}

\title{Precise determination of the strong coupling constant in
$N_f=2+1$ lattice QCD with the Schr\"odinger functional scheme}

\author{S.~Aoki$^{1,2}$, K.~-I.~Ishikawa$^3$, N.~Ishizuka$^{1,4}$,
T.~Izubuchi$^2$, D.~Kadoh$^5$, K.~Kanaya$^1$, Y.~Kuramashi$^{1,4}$,
K.~Murano$^{1}$, Y.~Namekawa$^4$, M.~Okawa$^3$, Y.~Taniguchi$^{1,4}$,
A.~Ukawa$^4$, N.~Ukita$^4$ and T.~Yoshi\'e$^{1,4}$\\
(PACS-CS collaboration)
}
\affiliation{
$^1$Graduate School of Pure and Applied Sciences,
University of Tsukuba,
Tsukuba, Ibaraki 305-8571, Japan\\
$^2$Riken BNL Research Center, Brookhaven National Laboratory, Upton,
New York 11973, USA\\
$^3$Graduate School of Science, Hiroshima University, Higashi-Hiroshima,
Hiroshima 739-8526, Japan\\
$^4$Center for Computational Physics,
University of Tsukuba,
Tsukuba, Ibaraki 305-8577, Japan\\
$^5$Theoretical Physics Laboratory,
The Institute of Physical and Chemical Research (RIKEN),
Wako, Saitama 351-0198, Japan 
}

\date{\today}

\begin{abstract}
 We present an evaluation of the running coupling constant 
 for $N_f=2+1$ QCD.
 The Schr\"odinger functional scheme is used as the intermediate
 scheme to carry out non-perturbative running from the low energy region,
 where physical scale is introduced, to deep in the high energy perturbative
 region, where conversion to the ${\ovl{\rm MS}}$ scheme is safely performed. 
 Possible systematic errors due to the use of perturbation theory 
 occur only  in the conversion from three-flavor to four-flavor running 
 coupling constant near the charm mass threshold, where higher order 
 terms beyond 5th order in the $\beta$ function may not be negligible. 

 For numerical simulations we adopted  Iwasaki gauge action and
 non-perturbatively improved Wilson fermion action with the clover term.
 Seven renormalization scales are used to cover from low to high energy
 region and three lattice spacings to take the continuum limit at each
 scale.

 A physical scale is introduced from the previous $N_f=2+1$ simulation of
 the CP-PACS/JL-QCD collaboration \cite{Ishikawa:2007nn}, which covered 
 the up-down quark mass range heavier than $m_\pi\sim 500$~MeV.
 Our final result is
$\alpha_{\ovl{\rm MS}}(M_Z)=0.12047(81)(48)(^{+0}_{-173})$ and
$\Lambda_{\ovl{\rm MS}}^{(N_f=5)}=239(10)(6)(^{+0}_{-22})$ MeV .

\end{abstract}

\maketitle

\section{Introduction}

The strong coupling constant and quark masses constitute the 
fundamental parameters of the Standard Model.
It is an important task of lattice QCD to determine these
parameters using inputs at low energy scales such as 
hadron masses, meson decay constants and quark potential quantities. 
The results can be compared with independent determinations from high
energy experiments, which  should provide a firm evidence of the single scale 
nature of QCD. 

In the course of evaluating these fundamental parameters we need the process
of renormalization in some scheme.
The ${\ovl{\rm MS}}$ scheme is one of the most popular schemes, and 
hence one would like to evaluate the running coupling constant 
through input of low energy quantities on the lattice and convert it to
the ${\ovl{\rm MS}}$ scheme.
A difficulty in this process is that the conversion is given only in a
perturbative expansion, and should be performed at high energy scales much 
larger than the QCD scale.
At the same time the renormalization scale $\mu$ should be kept much less
than the lattice spacing to reduce lattice artifacts, namely we require 
\begin{eqnarray}
 \Lambda_{\rm QCD}\ll\mu\ll\frac{1}{a}
\label{eqn:window}
\end{eqnarray}
A practical difficulty of satisfying these inequalities in numerical 
simulations is called the window problem. 

One of the widely used definitions of the renormalized coupling on the
lattice is to employ quantities related to the heavy quark potential
\cite{Lepage:1992xa,Schroder:1998vy}, which are easy to measure
accurately.
Choosing small size Wilson loops, the running coupling constant is 
extracted from their perturbative expansion with the renormalization scale 
set to $\mu\simeq1/a$. 
This conflicts with the window, and as a consequence
lattice artifacts are intrinsically included in the 
perturbative expansion coefficient in terms of the $\ovl{\rm MS}$
coupling.
The coefficients tend to explode, which is partly cured by the tadpole
improvement \cite{Lepage:1992xa} and by combining with $O(a)$ improved
actions like the staggered fermion action.
This definition of the coupling constant has been employed 
for $N_f=0$ \cite{ElKhadra:1992vn,Lepage:1993xy}, $N_f=2$ 
\cite{Aoki:1994pc,Gockeler:2004ad} and $N_f=2+1$ 
\cite{Davies:2003ik,Mason:2005zx,Davies:2008sw,Davies:2008nq,Maltman:2008vj}
flavor cases (for a review see Ref.~\cite{Weisz:1995yz,Rakow:2004vj}).
Recently developed methods using moments of charm quark current-current
correlator \cite{Allison:2008xk} or vacuum polarization function
\cite{Shintani:2008ga} are also not free from the window problem when
applying the perturbative expansion while reducing the lattice artifact.

The Schr\"odinger functional (SF) scheme
\cite{Luscher:1992an,Luscher:1993gh,Sint:1993un,Sint:2000vc,DellaMorte:2004bc}
is designed to resolve the window problem.
It has an advantage that systematic errors can be unambiguously
controlled.
A unique renormalization scale is introduced through the box size $L$.
A wide range of renormalization scales can be covered by the step
scaling function (SSF) technique, which exempts us of the requirement to
satisfy the condition \eqn{eqn:window} in a single simulation.
This matches our goal to obtain the coupling constant in the $\ovl{\rm MS}$
scheme and make comparisons with high energy inputs.
The SF scheme has been applied for evaluation of the QCD coupling for
$N_f=0$ \cite{Luscher:1993gh} and $N_f=2$ \cite{DellaMorte:2004bc}.

In the SF scheme we start with the evaluation of the running coupling constant for
a variety of the bare coupling constant $\beta$ and box sizes, which covers the
strong coupling region corresponding to the energy scale $\mu\sim500$ MeV
and the weak coupling region around $\mu\sim40$ GeV.
At low energy scales we expect the strange quark contribution to be important 
in addition to those of the up and down quarks. 
Thus the aim of the present paper is to go one step further than those of 
Refs.~\cite{Luscher:1993gh, DellaMorte:2004bc} and evaluate the 
strong coupling constant in $N_f=2+1$ QCD.  
For setting the physical scale we employ a recent large-scale $N_f=2+1$ 
lattice QCD simulation employing non-perturbatively O(a) improved Wilson 
quark action; the work of CP-PACS/JL-QCD Collaboration with relatively heavy 
pion mass with $m_\pi\sim 500$~MeV \cite{Ishikawa:2007nn}.

\reseteqnum
\section{Schr\"odinger functional formalism and action}

The Schr\"odinger functional is defined on a finite box
of size $L^3\times T$ with the Dirichlet boundary condition at the temporal
boundary.
For QCD the Dirichlet boundary condition is set for the spatial component of
the gauge link
\begin{eqnarray}
&&
U_k(x)|_{x_0=0}=\exp\left(a C_k\right),\quad
U_k(x)|_{x_0=T}=\exp\left(a C_k'\right),
\\&&
C_k=
\frac{i}{L}\pmatrix{\phi_1\cr&\phi_2\cr&&\phi_3\cr},\quad
C_k^{'}=
\frac{i}{L}\pmatrix{\phi_1^{'}\cr&\phi_2^{'}\cr&&\phi_3^{'}\cr}
\label{eqn:boundary}
\end{eqnarray}
and for the quark fields
\begin{eqnarray}
\psi(x)|_{x_0=0}=\psi(x)|_{x_0=T}=0,\quad
\bpsi(x)|_{x_0=0}=\bpsi(x)|_{x_0=T}=0.
\end{eqnarray}
Under a mild assumption it is proven that the tree level gauge
effective action has a global minimum around a background field
\cite{Luscher:1992an}
\begin{eqnarray}
&&
V_\mu(x)=\exp\left(aB_\mu(x)\right),
\\&&
B_0=0,\quad
B_k=\frac{1}{T}\left(x^0C_k'+(T-x^0)C_k\right),
\end{eqnarray}
which is uniquely given by the boundary fields \eqn{eqn:boundary}.
The fermionic mode is shown to have a mass gap
\cite{Sint:1993un}, so that we are able to define a mass independent
scheme directly in the chiral limit.

In this paper we adopt the same set up (scheme) as the Alpha
collaboration \cite{Luscher:1993gh,DellaMorte:2004bc} for the boundary link
\eqn{eqn:boundary}
\begin{eqnarray}
&&
\pmatrix{\phi_1\cr&\phi_2\cr&&\phi_3\cr}=
\eta\pmatrix{\omega_1\cr&\omega_2\cr&&\omega_3\cr}
+\pmatrix{-\frac{\pi}{3}\cr&0\cr&&\frac{\pi}{3}\cr},
\label{eqn:BC-ck}
\\&&
\pmatrix{\phi'_1\cr&\phi'_2\cr&&\phi'_3\cr}
=-\eta\pmatrix{\omega_1\cr&\omega_3\cr&&\omega_2\cr}
+\pmatrix{-\pi\cr&\frac{1}{3}\pi\cr&&\frac{2}{3}\pi\cr},
\label{eqn:BC-ckp}
\\&&
\pmatrix{\omega_1\cr&\omega_2\cr&&\omega_3\cr}
=\pmatrix{1\cr&-\frac{1}{2}\cr&&-\frac{1}{2}\cr}
+\nu\pmatrix{0\cr&1\cr&&-1\cr}. 
\end{eqnarray}
The parameter $\eta$ is used to define the renormalized coupling constant 
from the derivative of the effective action and is set to zero in the
action after taking derivative with respect to it.
The parameter $\nu$ may be used to define another renormalized quantity, 
but we set it to zero when evaluating the coupling constant.
We employ the choice $T=L$ so that the renormalization scale is given
by the box size $L$.

We adopt the renormalization group improved gauge action of Iwasaki given by 
\begin{eqnarray}
S_g=
 \frac{\beta}{N}\sum_{{ C}\in{ S}_0}W_0({ C},g_0^2)
{\rm Re\ }\tr\left(1-P({ C})\right)
+\frac{\beta}{N}\sum_{{ C}\in{ S}_1}W_1({ C},g_0^2)
{\rm Re\ }\tr\left(1-R({ C})\right),
\end{eqnarray}
where $S_0$ and $S_1$ are the sets of oriented plaquettes and rectangles.
The weight factor $W_{0/1}$ is chosen to cancel the $O(a)$
contribution from the boundary
according to \cite{Takeda:2003he,Takeda:2004xh}.
\begin{eqnarray}
&&
W_0({ C},g^2_0) = 
\left\{
 \begin{array}{ll}
  c_0 c^P_{\rm{t}}(g^2_0) 
   &    \mbox{Set of temporal plaquettes that just touch} 
   \\
  &
   \mbox{one of the boundaries, }
   \\      
  c_0        
   & \mbox{otherwise, } 
 \end{array}
\right.
\label{eqn:W0}
\\&&
W_1({ C},g^2_0) = 
\left\{
 \begin{array}{ll}
  c_1 c^R_{\rm{t}}(g^2_0) 
   &    \mbox{Set of temporal rectangles that have exactly} 
   \\
  &
   \mbox{two links on a boundary, }
   \\     
  c_1        
   &    \mbox{otherwise, }     
 \end{array}
\right.
\label{eqn:W1}
\end{eqnarray}
The bulk coefficients are set to $c_1 = -0.331$, $c_0 + 8c_1=1$.
The boundary improvement coefficients are set to the tree-level values
$c^P_{\rm{t}}=1$ and $c^R_{\rm{t}}=3/2$; it is empirically known 
that they give better scaling behavior than the one-loop values for the
$N_f=0$ \cite{Takeda:2004xh} and $N_f=2$ case
\cite{Murano:2009qi}.

We used the improved Wilson fermion action with clover term
\begin{eqnarray}
&&
S_f[U,\psi,\bpsi]
=a^4\sum_{x}\bpsi\left(D_W+m_0\right)\psi,
\\&&
D_W=\frac{1}{2}\left(\gamma_\mu\left(\nabla_\mu+\nabla_\mu^*\right)
-a\nabla_\mu^*\nabla_\mu\right)
-c_{\rm SW}\frac{1}{4}\sigma_{\mu\nu}P_{\mu\nu}.
\end{eqnarray}
The improvement coefficient $c_{\rm SW}$ is given non-perturbatively in
a polynomial form for $N_f=3$ QCD with the Iwasaki action by 
\cite{Aoki:2005et}
\begin{eqnarray}
c_{\rm SW}\left(g_0\right)=1+0.113g_0^2+0.0209(72)g_0^4+0.0047(27)g_0^6,
\end{eqnarray}
which covers $1.9\le\beta\le12.0$.
Although $O(a)$ effects in the bulk is canceled by the clover term, 
there are $O(a)$ contributions from the boundary for the
SF formalism and we need to add the boundary term to cancel it, 
\begin{eqnarray}
S_{{ O}(a)}&=&
a^3\sum_{\vec{x}}\left(\wt{c}_t-1\right)
\left(\bpsi(\vec{x},1)\psi(\vec{x},1)+\bpsi(\vec{x},T-1)\psi(\vec{x},T-1)
\right).
\end{eqnarray}
The coefficient is set to the one loop value given by \cite{Aoki:1998qd}
\begin{eqnarray}
\wt{c}_{t}=1-0.00881(28)g_0^2.
\end{eqnarray}
We employ the twisted periodic boundary condition in the three spatial
directions,
\begin{eqnarray}
\psi(x+L\hat{k})=e^{i\theta}\psi(x),\quad
\bpsi(x+L\hat{k})=e^{-i\theta}\bpsi(x)
\end{eqnarray}
with the same $\theta={\pi}/{5}$ for all spatial directions, as was used by the Alpha collaboration
\cite{Luscher:1993gh,DellaMorte:2004bc}.

The renormalized gauge coupling in the SF scheme is defined from
the effective action $\Gamma[V_\mu]$ at the global minimum.
For numerical simulation we take the derivative in terms of the parameter
$\eta$ introduced in the background field $\phi_i$ and define the SF
coupling constant as \cite{Luscher:1993gh}
\begin{eqnarray}
\frac{1}{\ovl{g}^2(L)}=
\frac{1}{k}\left.\frac{\p\Gamma[V_\mu]}{\p\eta}\right|_{\eta=0},
\end{eqnarray}
where
\begin{eqnarray}
k=12\left(\frac{L}{a}\right)^2
\left(c_0\left(\sin\xi+\sin2\xi\right)
+4c_1\left(\sin2\xi+\sin4\xi\right)\right)
,\quad
\xi=\frac{1}{3}\pi\left(\frac{a^2}{TL}\right)
\end{eqnarray}
is a normalization coefficient evaluated at tree level.

\reseteqnum
\section{Our strategy}
\label{sec:strategy}

Our goal is to derive the renormalization group invariant
(RGI) scale $\Lambda_{\rm QCD}$ in physical units and evaluate the
running coupling constant $\alpha_s(M_Z)$ at high energy scale $\mu=M_Z$.
The RGI scale $\Lambda$ is scheme dependent and we employ the commonly
used definition for the SF scheme,
\begin{eqnarray}
\Lambda_{\rm SF}=
\frac{1}{L}\left(b_0\ovl{g}(L)\right)^{-\frac{b_1}{2b_0^2}}
\exp\left(-\frac{1}{2b_0\ovl{g}(L)}\right)
\exp\left(-\int_{0}^{\ovl{g}(L)}dg
\left(\frac{1}{\beta(g)}+\frac{1}{b_0g^3}-\frac{b_1}{b_0^2g}\right)\right),
\label{eqn:lambda}
\end{eqnarray}
where $\ovl{g}(L)$ is the SF renormalized coupling at the box scale 
$L$ and $\beta(g)$ is the renormalization group $\beta$ function
in the same scheme whose perturbative expansion
coefficients are given by \cite{Bode:1999sm}
\begin{eqnarray}
&&
\beta(g)=-g^3\left(b_0+b_1g^2+b_2g^4+\cdots\right),
\\&&
b_0=\frac{1}{(4\pi)^2}\left(11-\frac{2}{3}N_f\right),
\\&&
b_1=\frac{1}{(4\pi)^4}\left(102-\frac{38}{3}N_f\right),
\\&&
b_2=\frac{1}{(4\pi)^3}\left(0.483(7)-0.275(5)N_f+0.0361(5)N_f^2-0.00175(1)N_f^3
\right).
\end{eqnarray}

The derivation of the RGI scale for the SF scheme proceeds in the 
following steps \cite{Luscher:1993gh}:
\begin{enumerate}
\item[(i)] We start by calculating the step scaling function (SSF)
$\Sigma(u,a/L)$ on the lattice at several box sizes and lattice
spacings.
The SSF gives the relation between the renormalized coupling constants when the
renormalization scale is changed by some factor, which is fixed to 2 
in this paper, 
\begin{eqnarray}
\Sigma\left(u,\frac{a}{L}\right)=\left.\ovl{g}^2(2L)\right|_{u=\ovl{g}^2(L)}.
\end{eqnarray}
The scale is given by the box size $L$, 
and $a/L$ represents the discretization error.
We take sufficient number of values for the coupling $u$ to cover 
low to high energy scales. 
Taking the continuum limit at each scale $u$
\begin{eqnarray}
\sigma(u)=\lim_{a/L\to0}\Sigma\left(u,\frac{a}{L}\right),
\end{eqnarray}
and performing a polynomial fit we obtain a non-perturbative
running of the coupling constant in the SF scheme for the scale change of 2. 

\item[(ii)] In the second step we define a reference scale $L_{\rm max}$
through a fixed value of the renormalized coupling constant 
$\ovl{g}^2(L_{\rm max})$.
The value of $\ovl{g}^2(L_{\rm max})$ is arbitrary as long as it is well
in low energy region to suppress lattice artifacts with
$a/L_{\rm max}\ll1$.
We then start from $L_{\rm max}$ and follow the non-perturbative RG flow
in the SF scheme into the high energy region.
A typical scale turns out to be $1/L_{\rm max}\sim0.5$ GeV in this paper
so that after $n\sim5$ iterations the scale $1/L=2^{n}/L_{\rm max}\sim16$ GeV
is already in the perturbative region where the difference between perturbative
and non-perturbative RG runnings is negligible.

\item[(iii)] Substituting $\ovl{g}^2(L)$ and $L=2^{-n}L_{\rm max}$
into the definition \eqn{eqn:lambda} and evaluating the integral
with three loops $\beta$-function in the SF scheme \cite{Bode:1999sm}
for the weak coupling region we obtain the RGI scale
$\Lambda_{\rm SF}L_{\rm max}$ in terms of the reference scale.

\item[(iv)] In the last step we need some physical input measured in an
independent large scale simulation at some lattice spacing $a$ 
to quote $L_{\rm max}$ in physical units. 
The requirement for the lattice spacing and the reference scale is that
the magnitude of lattice artifacts $a/L_{\rm max}$ should be kept small.
In this paper we employ hadron masses for physical input and use the
lattice spacing determined from them in physical units as the 
intermediate scale.
We then obtain the RGI scale $\Lambda_{\rm SF}$ in physical units. 
The transformation into the $\msbar$ scheme is given exactly at one-loop order 
{\it via}
\begin{eqnarray}
\Lambda_{\ovl{\rm MS}}=2.61192\Lambda_{\rm SF}
\label{eqn:sf2ms}
\end{eqnarray}
for three flavors.
\end{enumerate}

The RGI scale $\Lambda_{\ovl{\rm MS}}$ measured so far is
for three flavors ($\Lambda_{\ovl{\rm MS}}^{(3)}$).
In order to evaluate the coupling constant $\alpha_s(M_Z)$ at high energy we need
to change the number of flavors at charm and bottom quark mass thresholds, 
obtaining $\Lambda_{\ovl{\rm MS}}^{(5)}$ for five flavors.
For this purpose we used the matching formula near mass thresholds
for the $\ovl{\rm MS}$ scheme at three-loop order in
Refs.~\cite{Larin:1994va,Chetyrkin:1997sg,Chetyrkin:1997un}.
The evaluation of $\alpha_s(M_Z)$ will proceed in the following steps
in this paper.
\begin{enumerate}
 \item[(i)] Introduce the physical scale through hadron masses and evaluate
	    $L_{\rm max}$ in units of GeV.
 \item[(ii)] Perform the non-perturbative step scaling $n=5$ times and reach
	     deep into the perturbative region $q\sim16$ GeV.
 \item[(iii)] Change the scheme to $\ovl{\rm MS}$ according to the two-loop
	      relation \cite{Bode:1999sm}
\begin{eqnarray}
&&
\alpha_{\ovl{\rm MS}}(sq)=
\alpha_{\rm SF}(q)+c_1(s)\alpha_{\rm SF}^2(q)+c_2(s)\alpha_{\rm SF}^3(q)
+\cdots,
\label{eqn:sf2msa}
\\&&
c_1(s)=-8\pi b_0\ln(s)+1.255621(2)+0.0398629(2)N_f,
\\&&
c_2(s)=c_1(s)^2-32\pi^2b_1\ln(s) +1.197(10)+0.140(6)N_f-0.0330(2){N_f}^2.
\nn\\
\end{eqnarray}
	      We may set the scale boost factor $s=2.61192$ so that
	      $c_1(s)=0$.
	      A systematic error due to higher loops correction is less
	      than $0.1$ \% and negligible here.
\item[(iv)] Running back to the charm quark mass threshold $\mu=m_c$ with
       the four loop $\beta$-function in the $\ovl{\rm MS}$ scheme we change
       the number of flavors to four using the three-loop matching
       formula \cite{Larin:1994va,Chetyrkin:1997sg,Chetyrkin:1997un}.
\begin{eqnarray}
\frac{\alpha^{(N_f-1)}(\mu)}{\pi}&=&
\frac{\alpha^{(N_f)}(\mu)}{\pi}F\left(\alpha^{(N_f)}(\mu),x\right),\quad
x=\ln\frac{M(\mu)^2}{\mu^2},
\label{eqn:nf2nf-1}
\\
F\left(\alpha,x\right)&=&
1+\sum_{k=1}^{3}F_k(x)\left(\frac{\alpha}{\pi}\right)^k,
\\
F_1(x)&=&\frac{1}{6}x,
\\
F_2(x)&=&F_1(x)^2+\frac{11}{24}x+\frac{11}{72},
\\
F_3(x)&=&\frac{564731}{124416}-\frac{82043}{27648}\zeta(3)
+\frac{955}{576}x+\frac{53}{576}x^2+\frac{1}{216}x^3
\nn\\&&
+\left(N_f-1\right)
\left(-\frac{2633}{31104}-\frac{67}{576}x-\frac{1}{36}x^2\right),
\end{eqnarray}
       where $M(\mu)$ is the $\overline{\rm MS}$ running mass of the heavy quark which decouples 
       at the threshold.
       We shall set $\mu=M(M)$ and $x=0$ in this paper.
       Since the largest error may be introduced from
       the use of perturbation theory at $\mu=m_c(m_c)$,
       we estimate the systematic error of this perturbative
       matching,  by comparing the result with that from the two-loop
       matching relation\cite{Bernreuther:1981sg,Bernreuther:1983zp,Rodrigo:1993hc}.     
\item[(v)] Running to the bottom quark mass threshold $\mu=m_b(m_b)$ we obtain the
       running coupling constant for five flavors in the same manner.
\item[(vi)] Finally we change the scale to $\mu=M_Z(M_Z)$ with the four-loop
       $\beta$-function and find $\alpha_s(M_Z)$.
\item[(vii)] The RGI scale $\Lambda_{\ovl{\rm MS}}^{(5)}$ is given by
       substituting $\mu=M_Z(M_Z)=1/L$ and $\alpha_s(M_Z)$ in the definition
       \eqn{eqn:lambda} for five flavors in the $\ovl{\rm MS}$ scheme
	    with the four-loop $\beta(g)$.
\end{enumerate}

\reseteqnum
\section{Step scaling function}

We adopt seven renormalized coupling values to cover weak
($\ovl{g}^2=1.001$) to strong ($\ovl{g}^2=3.418$) coupling regions,
which approximately satisfy $\ovl{g}^2_{i+1}(L) = \ovl{g}^2_i(2L)$
($i=1,\cdots,6$).
For each coupling we use three boxes $L/a=4, 6, 8$ to take the
continuum limit.

The HMC algorithm is adopted for two flavors and the RHMC algorithm for the
third flavor, all of which are set to a common mass of zero.
We adopt the CPS++ code \cite{cps} and add some modification for the SF
formalism.
Simulations were carried out on a number of computers, the PC cluster Kaede, 
PACS-CS and T2K-tsukuba at University of Tsukuba, T2K-tokyo and SR11000 
at University of Tokyo and the PC cluster RSCC at RIKEN.

\begin{figure}
 \begin{center}
  \includegraphics[width=4.5cm]{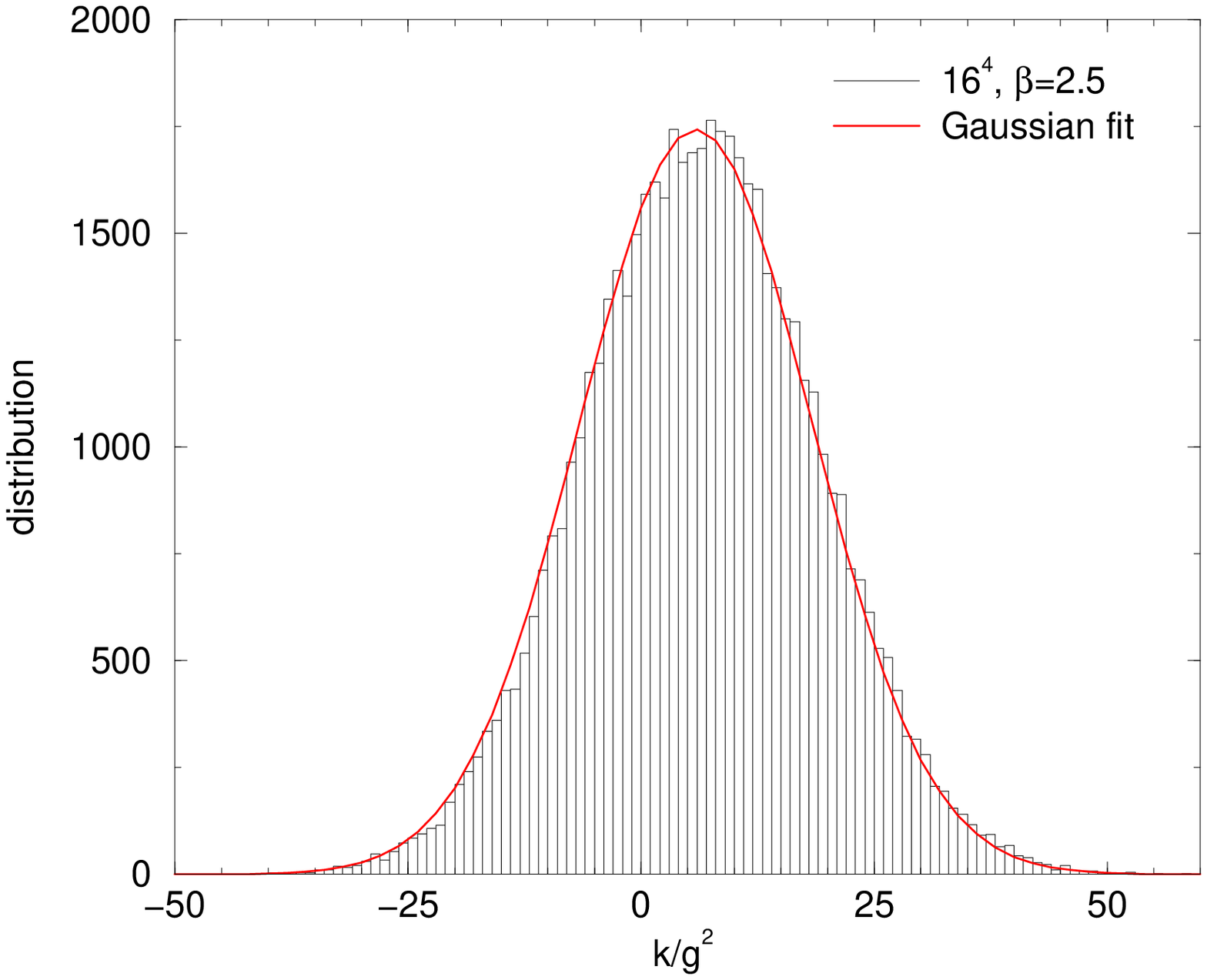}
  \includegraphics[width=4.5cm]{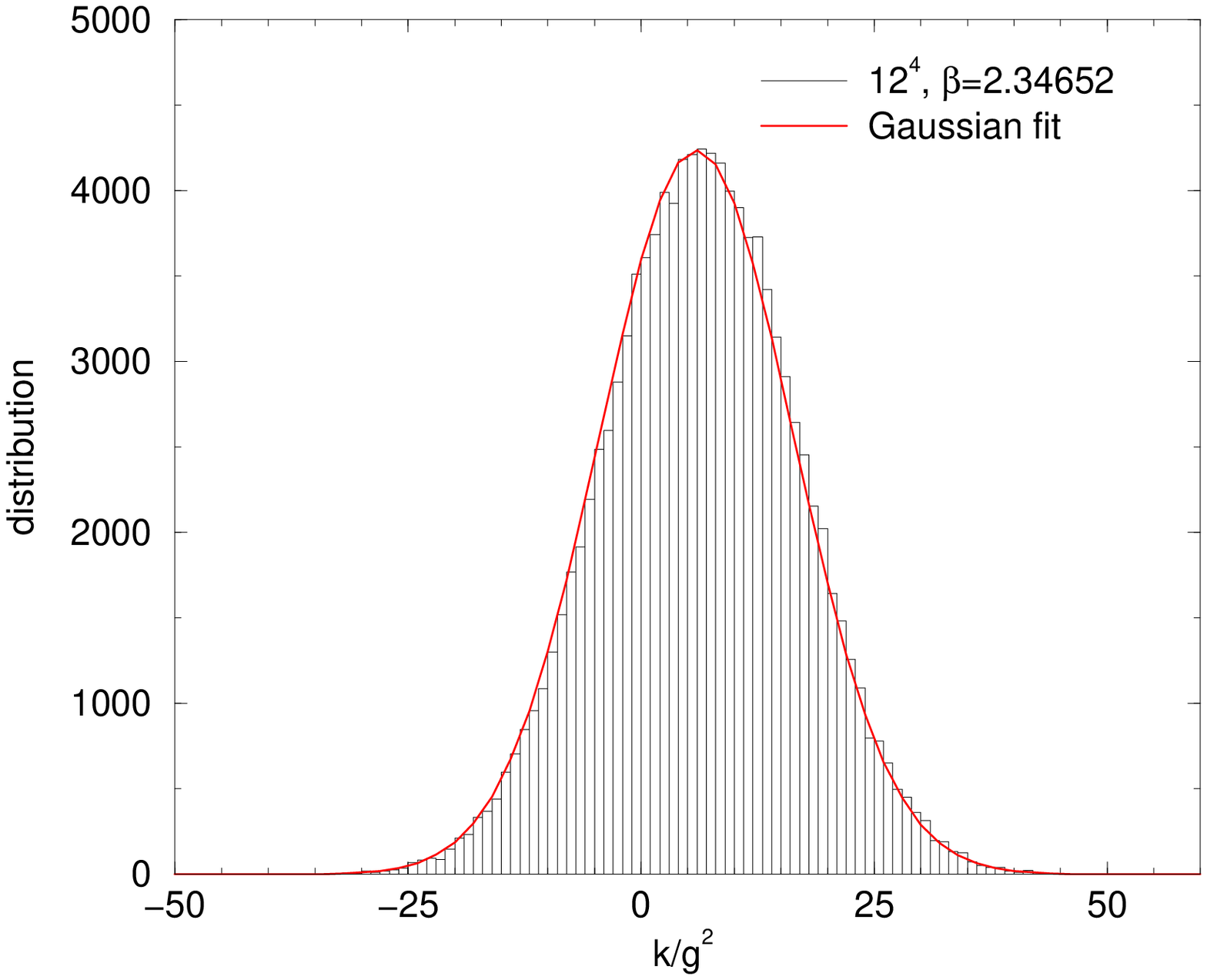}
  \includegraphics[width=4.5cm]{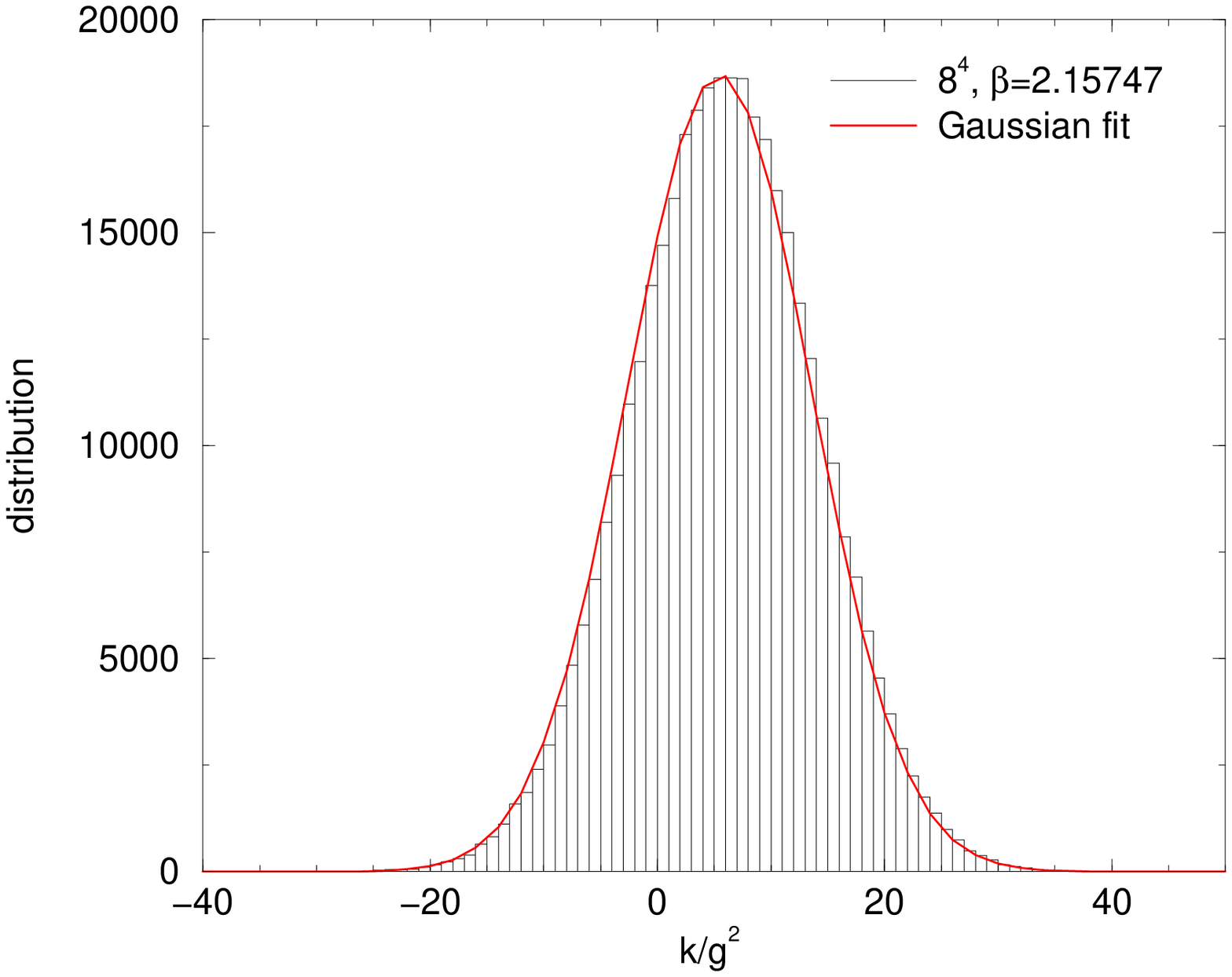}
  \caption{Distribution of inverse of the renormalized coupling at
  lowest energy scale given by $\ovl{g}^2(L)\sim5$, which corresponds to
  $L/a=16$, $\beta=2.5$ (left), $L/a=12$, $\beta=2.34652$ (middle) and
  $L/a=8$, $\beta=2.15743$ (right).
  Solid line is a fit in a Gaussian function.}
  \label{fig:distribution}
 \end{center}
\end{figure}

The distribution of the inverse of the coupling constant $1/\ovl{g}^2$
turned out to be a smooth Gaussian even at the lowest energy scale
\cite{Murano:2009qi} as plotted in Fig.~\ref{fig:distribution}.
This is contrary to the finding with the standard Wilson gauge action
\cite{Luscher:1993gh,DellaMorte:2004bc} and we need no re-weighting.

We start by tuning the value of $\beta$ and $\kappa$ to reproduce the
same renormalized coupling at each of the box sizes $4$, $6$, $8$
keeping the PCAC mass to zero.
Requirement for the renormalized couplings $\ovl{g}^2(L)$ is that their
values agree within one standard deviation for $L/a=4$, $6$, $8$.
The PCAC relation is defined in terms of the improved axial
current with non-perturbative improvement coefficient
\cite{Kaneko:2007wh}
\begin{eqnarray}
A_\mu^{\rm imp.}(x)=A_\mu(x)+c_A\p_\mu P(x), \quad
c_A(g_0^2)=-0.0038g_0^2\frac{1-0.195g_0^2}{1-0.279g_0^2}.
\label{eqn:axial-current}
\end{eqnarray}
The values of $(\beta,\kappa)$ are listed in Table \ref{tab:beta-kappa}
together with results for the renormalized coupling constant $\ovl{g}$
and the PCAC mass at the two scales $L$ and $2L$. 
Statistics of the runs are given in Table \ref{tab:beta-kappa-conf}.

The renormalized coupling $\ovl{g}^2(2L)$ at the scale $2L$ is corrected 
perturbatively in order to cancel the deviation of the PCAC mass from zero
at the scale $L$\cite{Sint:1995ch}
\begin{eqnarray}
&&
\left.\ovl{g}^2(2L)\right|_{\ovl{g}^2(L)=u,m=0}
=\left.\ovl{g}^2(2L)\right|_{\ovl{g}^2(L)=u,m(L)=m}
-\Phi(0)u^2mL,
\\&&
\Phi(0)=0.00957N_f.
\end{eqnarray}
The PCAC mass at the scale $L$ has been tuned such that the deviation $\Phi(0)u^2mL$ is
smaller than the typical statistical error.

The value of the renormalized coupling $\ovl{g}^2(L)$ at $L/a=8$ is used to define
$\ovl{g}^2(L)$ at scale $L$.
The deviation of $\ovl{g}^2(L)$ at $L/a=4,6$ from it  
is also corrected perturbatively at three-loop using 
\cite{Bode:1999sm}
\begin{eqnarray}
&&
\Sigma\left(u,\frac{a}{L}\right)
=\Sigma\left(\wt{u},\frac{a}{L}\right)
+\frac{\p\sigma_{\rm PT}^{(3)}}{\p u}(u-\wt{u}),
\\&&
\sigma_{\rm PT}^{(3)}(u)=u+s_0u^2+s_1u^3+s_2u^4.
\label{eq:ssf-pt}
\\&&
s_0=2b_0\ln2,
\label{eqn:s0}
\\&&
s_1=\left(2b_0\ln2\right)^2+2b_1\ln2,
\label{eqn:s1}
\\&&
s_2=\left(2b_0\ln2\right)^3+10b_0b_1\left(\ln2\right)^2+2b_2\ln2,
\label{eqn:s2}
\end{eqnarray}
where $b_n$ is the perturbative coefficient of the $\beta$-function in
the SF scheme.

We now consider the continuum extrapolation $a/L\to 0$ of the SSF. 
In perturbation theory the deviation of the lattice SSF from its continuum 
value is expressed as
\begin{eqnarray}
&&
\frac{\Sigma\left(u,{a}/{L}\right)-\sigma(u)}{\sigma(u)}
=\delta_1(a/L)u+\delta_2(a/L)u^2+\cdots,
\\&&
\delta_1(a/L)=\delta_{1G}(a/L)+N_f\delta_{1Q}(a/L).
\label{eqn:one-loop-deviation}
\end{eqnarray}
The one-loop coefficients $\delta_{1G/1Q}$ are given in Table \ref{tab:delta}
for the Iwasaki gauge action with the tree-level improved boundary coefficients
$c_t$ adopted for the present work 
for each box sizes \cite{Takeda:2003he,Takeda2009}.
As is seen from the table the values  of $\delta_{1Q/1G}$ are not small, 
and the deviation decreases only slowly with the volume $L/a$. 

\begin{figure}
 \begin{center}
  \includegraphics[width=4.5cm]{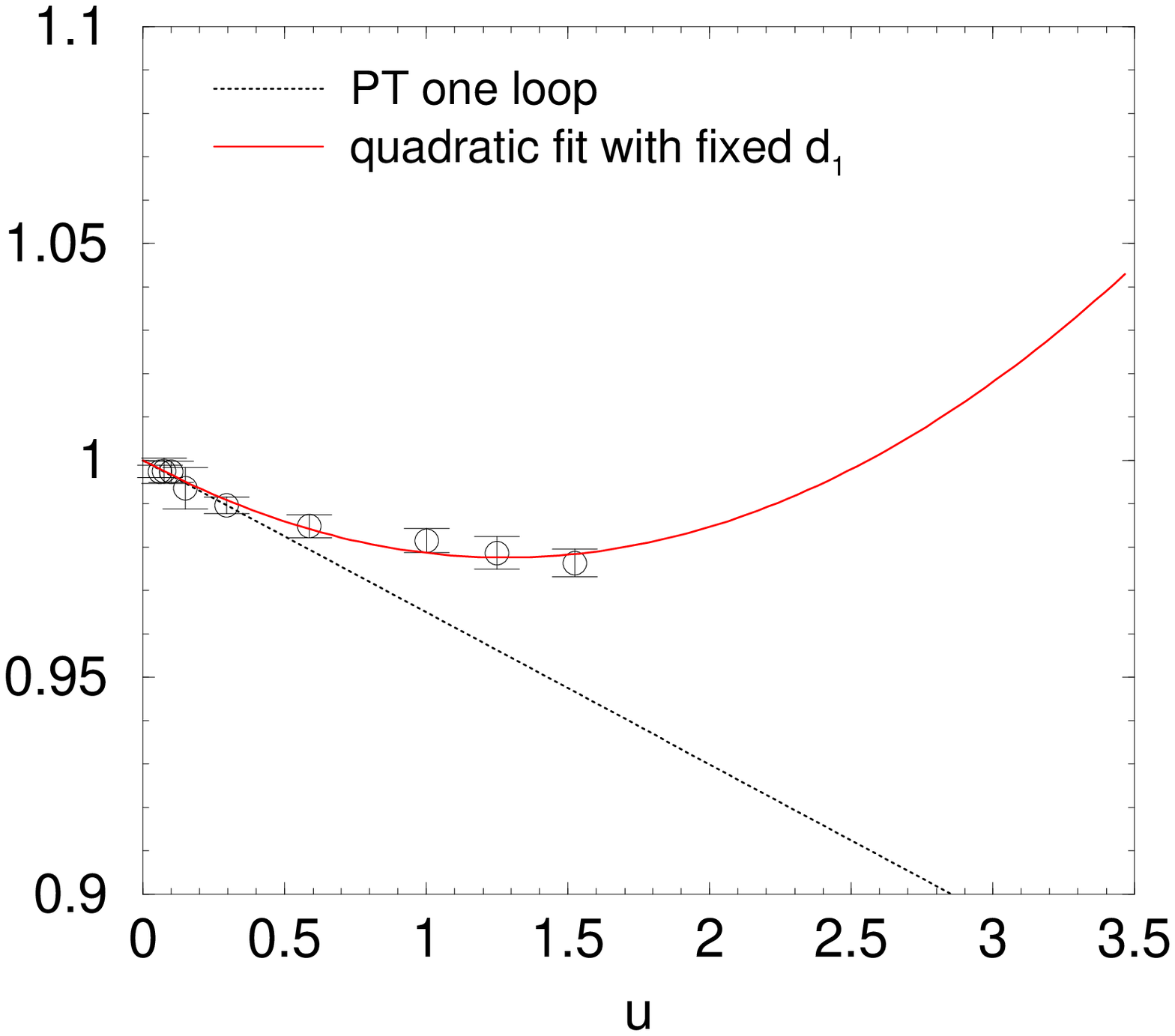}
  \includegraphics[width=4.5cm]{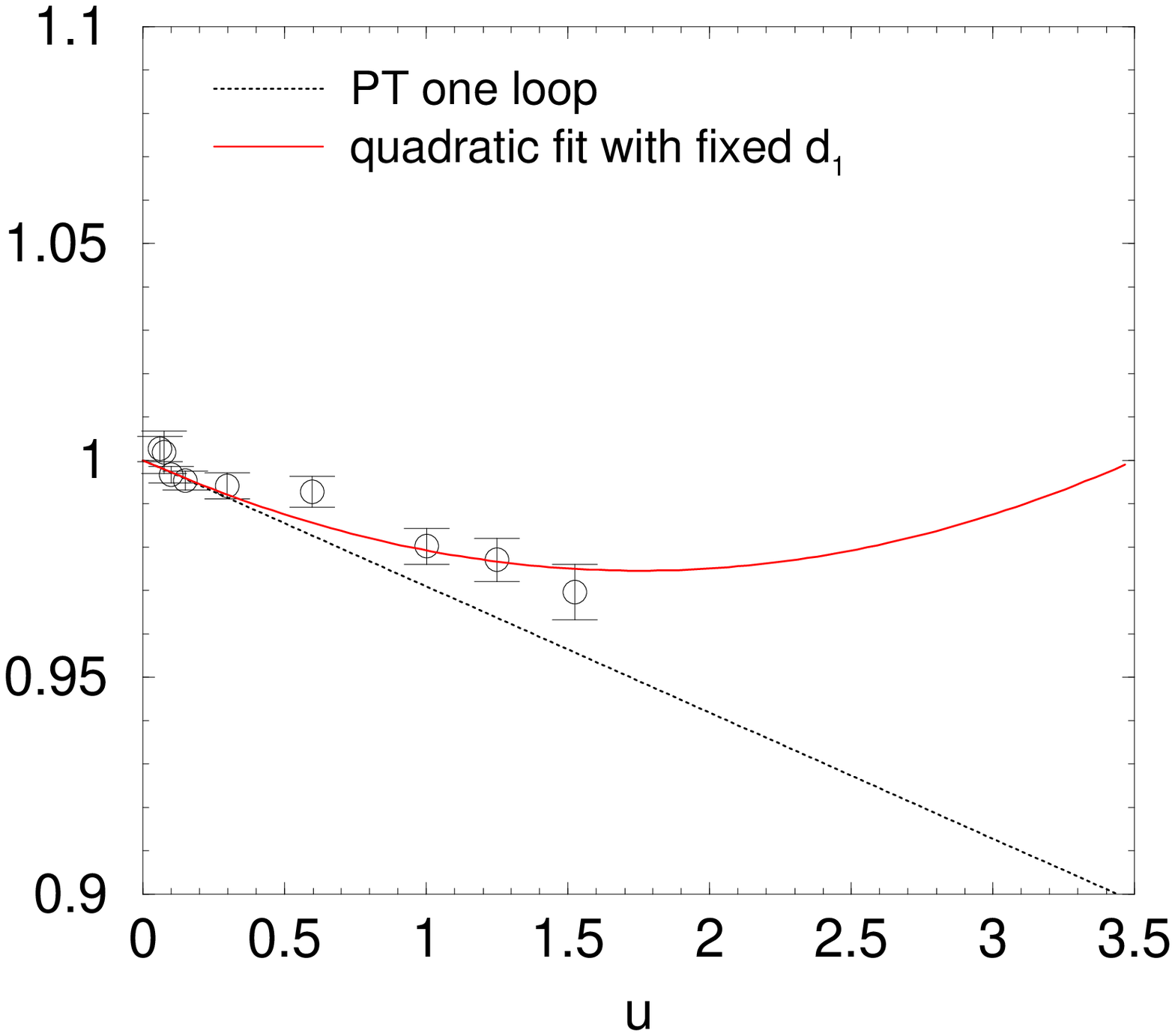}
  \includegraphics[width=4.5cm]{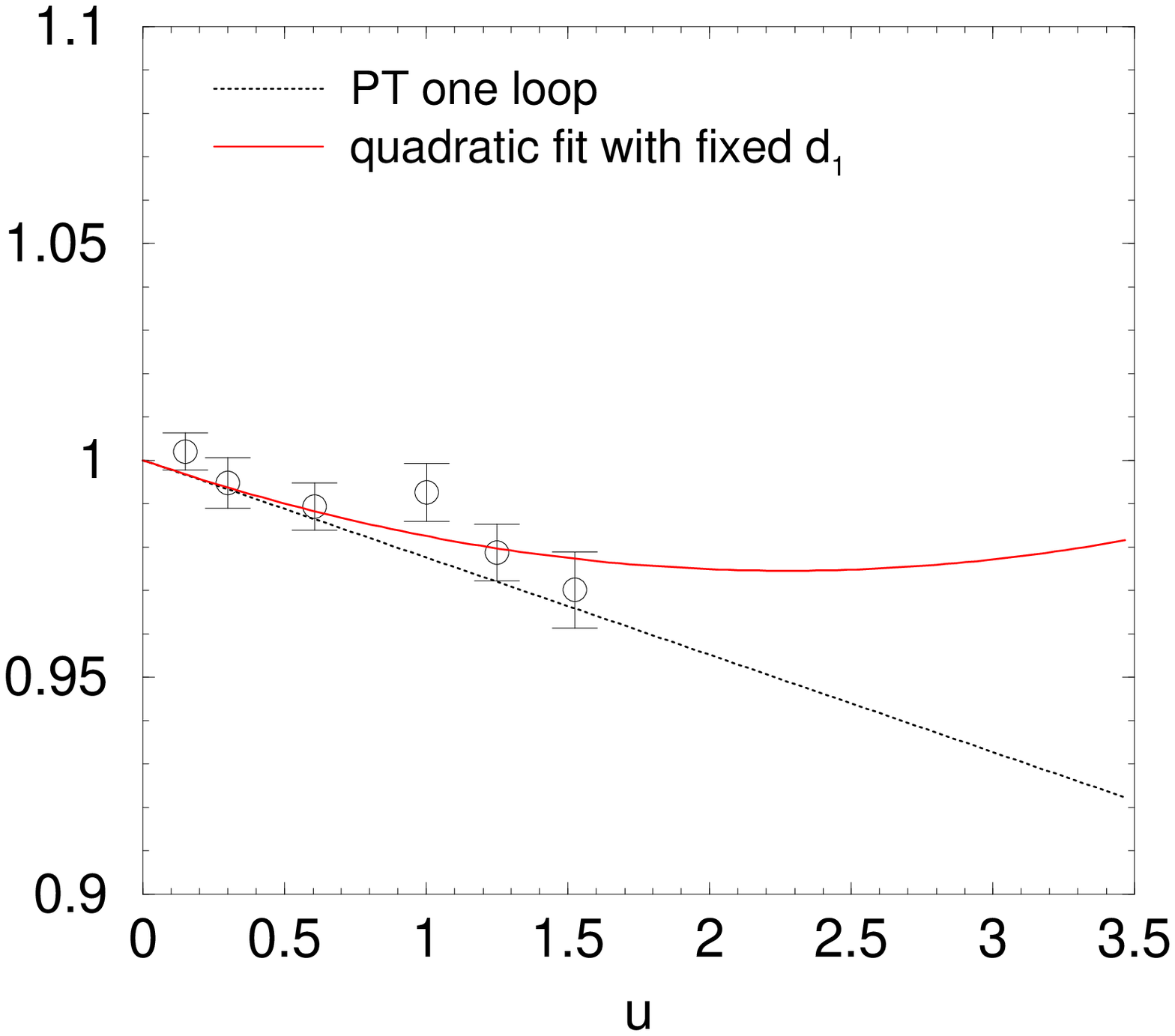}
  \caption{Polynomial fit of discrepancy
  $\Sigma\left(u,{a}/{L}\right)/\sigma_{\rm PT}^{(3)}(u)$ at high
  $\beta\gtrsim4$.
  The fit is given for each lattice spacings $a/L=1/4$ (left), $a/L=1/6$
  (middle) and $a/L=1/8$ (right).
  Black dotted line is a perturbative one loop behavior and red solid
  line is a quadratic fit with fixed $d_1$ to its one loop value.}
  \label{fig:ordera}
 \end{center}
\end{figure}

Instead of calculating the two-loop coefficients $\delta_{2Q/2G}$
perturbatively, which is a non-negligible task, we calculate SSF
directly by Monte-Carlo sampling at very weak coupling $\beta\ge10$.
The results are listed in Table \ref{tab:pt-beta-kappa}, where the
parameter is tuned only for $\kappa$ to reproduce $m_{\rm PCAC}=0$.
We define the deviation from the perturbative SSF
\begin{eqnarray}
\delta(u,a/L)=
\frac{\Sigma\left(u,{a}/{L}\right)-\sigma_{\rm PT}^{(3)}(u)}
{\sigma_{\rm PT}^{(3)}(u)},
\label{eqn:delta}
\end{eqnarray}
where $\sigma_{\rm PT}^{(3)}$ is the continuum SSF at three-loop order 
given by (\ref{eq:ssf-pt}). 
The deviation is fitted in a polynomial form for each $a/L$,
\begin{eqnarray}
1+\delta(u,a/L)=1+d_1(a/L)u+d_2(a/L)u^2.
\end{eqnarray}
We tried
a quadratic fit using data at $u\le1.524$ with fixing $d_1(a/L)$ to
its perturbative value $\delta_1(a/L)$, which is plotted in
Fig.~\ref{fig:ordera}.
We also plot perturbative one loop behavior for comparison.
As is seen from the figure the one loop line could reproduce the data
only at very high $\beta\ge10$ for $L/a=4,6$.
It may not be safe to adopt the one loop improvement for our data at
$u\ge1.0$.

The fit results for the coefficients are listed in table
\ref{tab:polynomial-fit}.
We observe that the higher-loop coefficient $d_2$ is not negligible
and contribute in opposite sign.
We notice that the fit result hardly changes even if we add one more
data at $u=1.840$.
Since the quadratic fit provides a reasonable description of data as
shown in Fig~\ref{fig:ordera}  we opt to cancel the $O(a)$ contribution
dividing out the SSF by the quadratic fit
according to 
\begin{eqnarray}
\Sigma^{\rm (2)}\left(u,\frac{a}{L}\right)=
\frac{\Sigma\left(u,{a}/{L}\right)}{1+\delta_1(a/L)u+d_2(a/L)u^2}.
\label{eqn:pt-imp}
\end{eqnarray}

\begin{figure}
 \begin{center}
  \includegraphics[width=8cm]{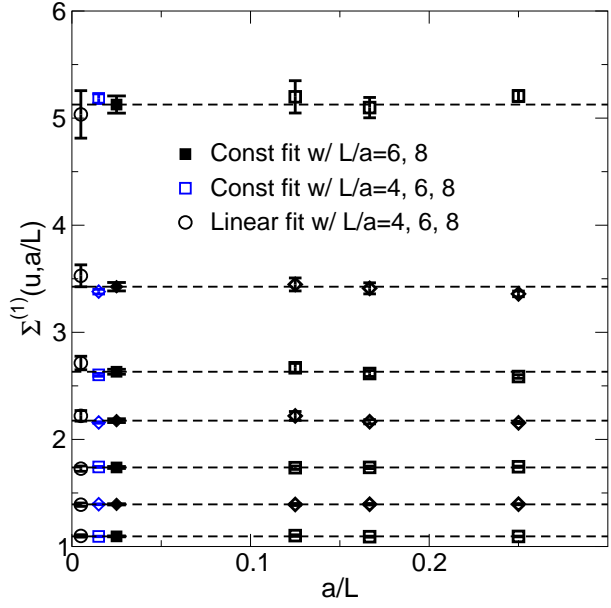}
  \caption{The SSF on the lattice with its continuum extrapolation at
  each renormalization scale.
  }
  \label{fig:SSF}
 \end{center}
\end{figure}

Now we have the values of the $O(a)$ improved SSF's in the chiral limit for
three lattice spacings at each of the 7 renormalization scale given by
$u$, which are listed in table \ref{tab:ssf}.
Scaling behavior of the SSF is plotted in Fig.~\ref{fig:SSF}.  
Almost no scaling violation is found.
We performed three types of continuum extrapolation:
a constant extrapolation with the finest two (filled symbols) or all three data
points (open symbols), or a linear extrapolation with all three data
points (open circles).
As is shown in the figure they are consistent with each other.
Since the scaling behavior is very good for the finest two lattice spacings
we employed the constant fit with these two data point to find our continuum
value, which is also listed in Table \ref{tab:ssf}.

\begin{figure}
 \begin{center}
  \includegraphics[width=8cm]{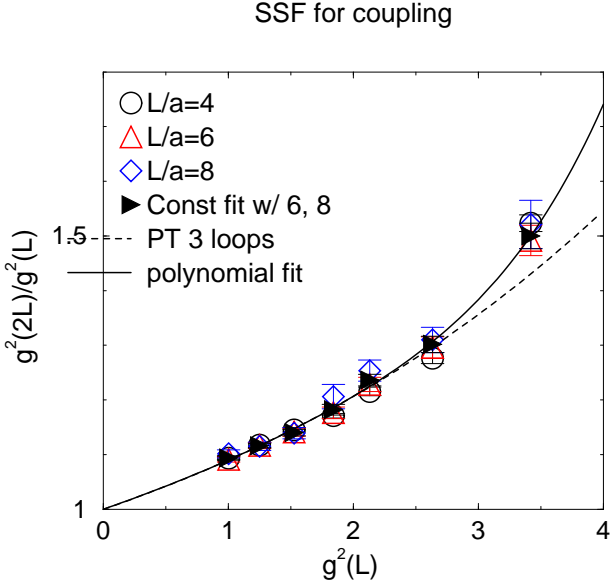}
  \caption{RG flow of the SSF divided by the coupling $\ovl{g}^2(L)$.
  Dotted line is three loops perturbative running.
  Solid line is a polynomial fit of the SSF.
  }
  \label{fig:SSF-fit}
 \end{center}
\end{figure}

The RG running of the continuum SSF is plotted in
Fig.~\ref{fig:SSF-fit}.
We divide the SSF with the coupling $\ovl{g}^2(L)$ to obtain a better
resolution in this figure.
A polynomial fit of the continuum SSF to sixth order
fixing the first and second coefficients $s_0$ and $s_1$
to their perturbative values \eqn{eqn:s0}, \eqn{eqn:s1} yields 
\begin{eqnarray}
&&
\sigma(u)=u+s_0u^2+s_1u^3+s_2u^4+s_3u^5+s_4u^6,
\label{eqn:poly}
\\&&
s_2=0.002265,\quad
s_3=-0.00158,\quad
s_4=0.000516.
\end{eqnarray}
The fitting function is also plotted (solid line) together with the
three loop perturbative running (dashed line).

\subsection{Non-perturbative $\beta$-function}

From the polynomial form of the SSF we derive the non-perturbative
$\beta$-function for $N_f=3$ QCD.
Starting from definition of the $\beta$-function
\begin{eqnarray}
-L\frac{\p u(L)}{\p L}=2\sqrt{u}\beta(\sqrt{u}),\quad
u=\ovl{g}^2(L)
\end{eqnarray}
the value of the $\beta$-function at stronger coupling (lower scale) is
given by recursively solving the relation
\begin{eqnarray}
\beta\left(\sqrt{\sigma(u)}\right)=
\beta(\sqrt{u})\sqrt{\frac{u}{\sigma(u)}}\frac{\p \sigma(u)}{\p u}.
\end{eqnarray}
The input is the three loops perturbative value at $u=0.9381$, which is
deep in the perturbative region.

For the non-perturbative SSF we adopt a slightly different fitting form
in order to reduce the error propagation.
We performed a polynomial fit by fixing the first to third coefficients
$s_0$, $s_1$ and $s_2$ to their perturbative values \eqn{eqn:s0},
\eqn{eqn:s1}, \eqn{eqn:s2}
\begin{eqnarray}
&&
\sigma(u)=u+s_0u^2+s_1u^3+s_2u^4+s_3u^5+s_4u^6,
\\&&
s_3=-0.000673,\quad
s_4=0.0003434.
\end{eqnarray}

\begin{figure}
 \begin{center}
  \includegraphics[width=8cm]{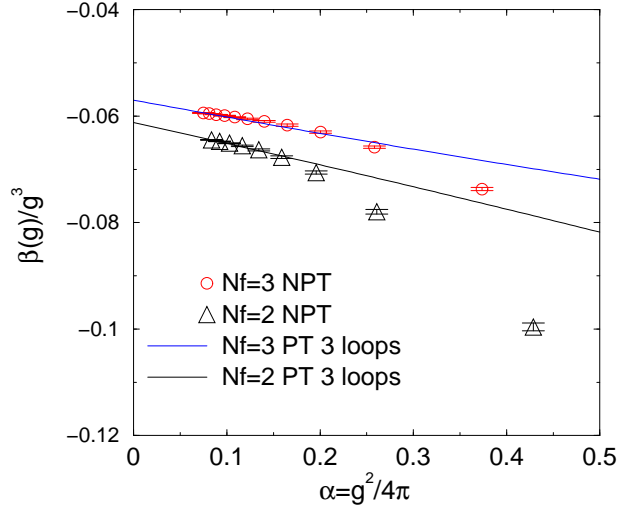}
  \caption{Non-perturbative $\beta$-function for $N_f=3$ and $2$ QCD.
  Solid lines are three loops perturbative running for comparison.
  Data for $N_f=2$ is reproduced from Ref.~\cite{DellaMorte:2004bc}.
  }
  \label{fig:npt-beta}
 \end{center}
\end{figure}

The resultant $\beta$-function is plotted in Fig.~\ref{fig:npt-beta}.
The $\beta$-function of $N_f=2$ QCD is reproduced from data of the Alpha
collaboration \cite{DellaMorte:2004bc} for comparison.
Note that the error is estimated by a propagation from those in the
continuum SSF's $\sigma(u)$.

\reseteqnum
\section{Introduction of physical scale}

CP-PACS and JLQCD Collaborations jointly performed an $N_f=2+1$
simulation with the $O(a)$ improved Wilson action and the Iwasaki gauge action, 
whose results have been recently published \cite{Ishikawa:2007nn}.
Three values of $\beta$, $1.83$, $1.90$ and $2.05$ were adopted to take the
continuum limit and the up-down quark mass covered a rather heavy region 
corresponding to $m_\pi/m_\rho=0.63-0.78$.

We adopt those results to introduce the physical scale into the present
work so that the reference scale $L_{\rm max}$ is translated into MeV units.
The Alpha Collaboration \cite{Luscher:1993gh,DellaMorte:2004bc} has
adopted the Sommer scale $r_0$ as a physical observable for this
purpose.
Since the Sommer scale is not a direct hadronic observable, we prefer 
to employ the hadron masses $m_\pi$, $m_K$, $m_\Omega$ as inputs 
and use the lattice spacing $a$ as an intermediate scale, 
which are listed in Table \ref{tab:lattice-spacing}.

We evaluate the renormalized coupling in the SF scheme
at the same $\beta=1.83$, $1.90$, $2.05$ in the chiral limit.
The reference scale $L_{\rm max}$ is given by the box size we adopt in
this evaluation. Note that this definition gives a different value of
$L_{\rm max}$ at different $\beta$.
The renormalized coupling $\ovl{g}^2(L_{\rm max})$ should
not exceed our maximal value $5.13$ of the SSF very much.
The value of the coupling constant at each $\beta$ are listed in Table
\ref{tab:lmax} together with the PCAC mass.
The hopping parameter $\kappa$ is tuned to reproduce $m_{\rm PCAC}=0$
except for the cases that the coupling constant apparently exceeds $5.13$.
We use the box size of $L/a=4$ for $\beta=1.83$ and $1.90$ to define
$L_{\rm max}$ and $L/a=4$, $6$ for $\beta=2.05$.

\reseteqnum
\section{RGI scale and the strong coupling constant at $M_Z$}

Starting from $u_{\rm max}=\ovl{g}^2(L_{\rm max})$ we iterate
the non-perturbative renormalization group flow five times according to
the polynomial fit \eqn{eqn:poly} and substitute the result 
$L=2^{-5}L_{\rm max}$ and $\ovl{g}(L)$ into \eqn{eqn:lambda} with
$\beta$-function for three flavors at three loops.
In this way we obtain $\Lambda_{\rm SF}^{(3)}L_{\rm max}$ for three
flavors.
Further non-perturbative step scaling with $n\ge6$ does not change the central
value of $\Lambda_{\rm SF}^{(3)}L_{\rm max}$.
The results are listed in Table \ref{tab:lambda}
together with $\Lambda_{\rm SF}^{(3)}$ in units of MeV and
$\Lambda_{\ovl{\rm MS}}^{(3)}$ given by \eqn{eqn:sf2ms}.

We derive the strong coupling constant $\alpha_s(M_Z)$ at high energy scale
$\mu=M_Z$ according to the procedure given in Sec.~\ref{sec:strategy}.
After reaching the scale $L=2^{-5}L_{\rm max}$ in the SF scheme, 
we transform to the $\ovl{\rm MS}$ scheme by the two-loop formula
\eqn{eqn:sf2msa} at $q=1/L$ with $s=\exp(c_1(1)/(8\pi b_0))$.
Then running back to the scale $\mu=m_c(m_c)$ with three-flavor
4-loop $\beta$-function the coupling constant is matched to that for four flavors at
three-loop order using \eqn{eqn:nf2nf-1}.
We repeat the same operation at the threshold $\mu=m_b(m_b)$ and obtain the
five flavor coupling constant.
We finally run to $\mu=M_Z$ with the four-loop $\beta$-function for
five flavors and find $\alpha_s(M_Z)$.
The QCD parameter $\Lambda_{\ovl{\rm MS}}^{(5)}$ is given by substituting 
$\mu=M_Z=1/L$ and $\alpha_s(M_Z)$ in \eqn{eqn:lambda} for the $\ovl{\rm MS}$ scheme with
4-loop $\beta(g)$.
The results are listed in Table \ref{tab:alphas}.
For an estimate of the systematic error due to perturbation theory, results 
using three- and two-loop formula in \eqn{eqn:nf2nf-1} are listed.
The error includes the statistical error of the renormalized couplings,
which is propagated into that of the SSF, in addition to the statistical error
of the lattice spacing.
The experimental errors of $m_c$, $m_b$ and $M_Z$ are also included.

\begin{figure}[htbp]
 \includegraphics[width=7cm]{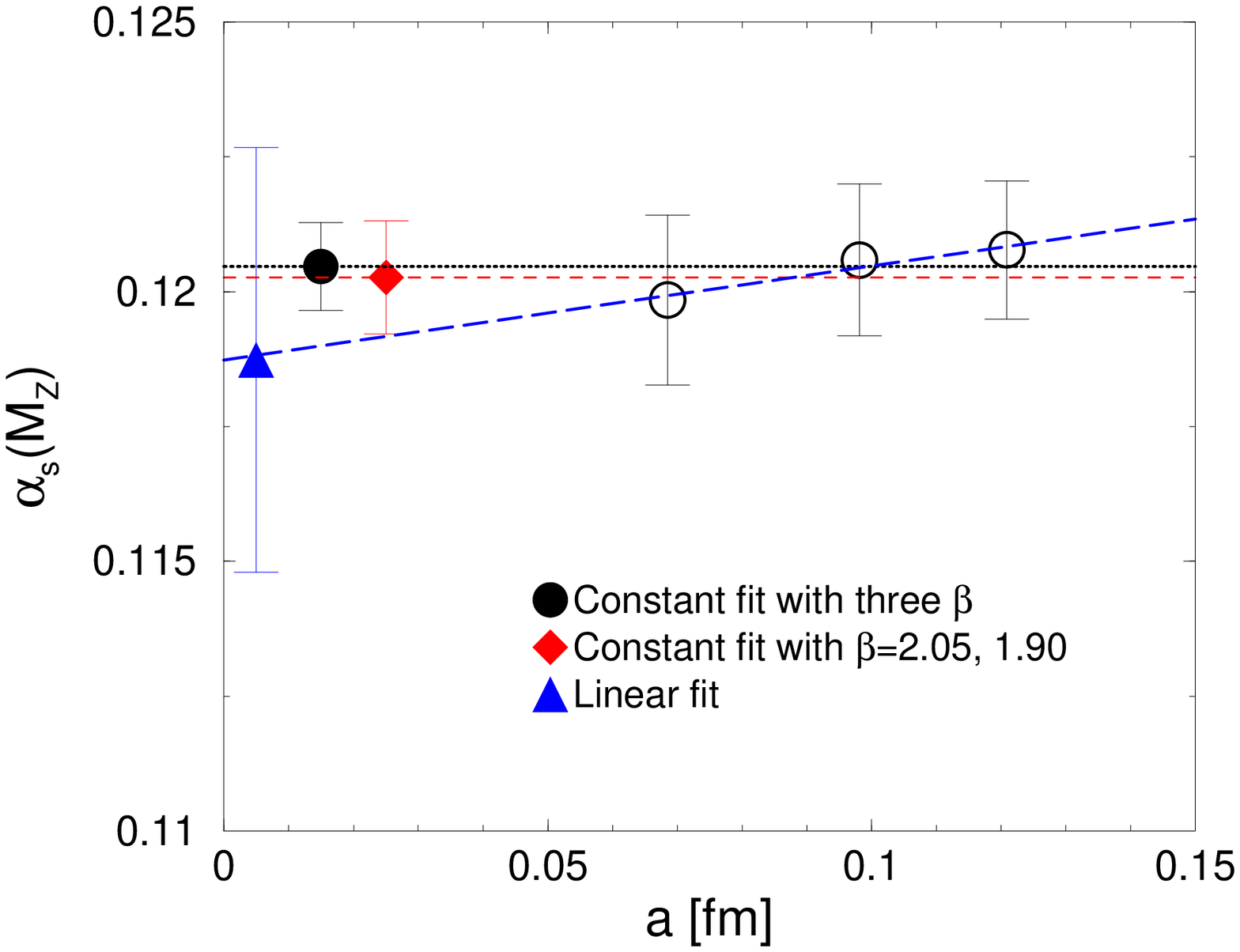}
 \includegraphics[width=7cm]{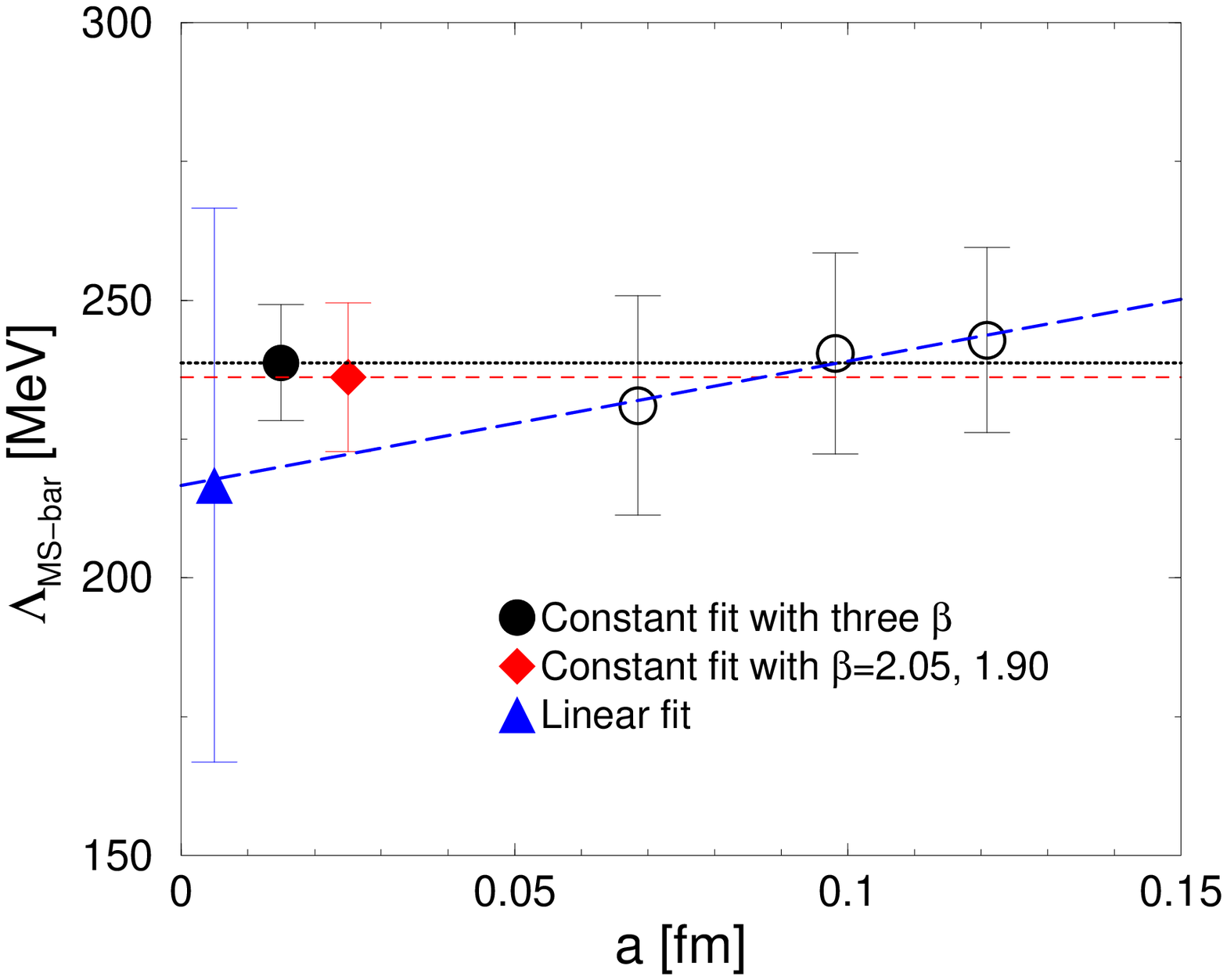}
 \caption{Scaling behavior of $\alpha_{\ovl{\rm MS}}(M_Z)$ (left) and 
 $\Lambda_{\ovl{\rm MS}}^{(5)}$ (right).
 We adopt $6^4$ data for $\beta=2.05$.
 Three types of continuum extrapolation is given; constant fit with
 three and two lattice spacings and linear fit.
}
\label{fig:alphamz}
\end{figure}

As the last step we take the continuum limit using the three lattice spacings
from Ref.~\cite{Ishikawa:2007nn}.
The scaling behavior of $\alpha_s(M_Z)$ and $\Lambda_{\ovl{\rm MS}}^{(5)}$
is plotted in Fig.~\ref{fig:alphamz}.
Since the results in the continuum limit do not depend on $L_{\rm max}$,
we adopt the result for $L=6$ as the central value for $\beta=2.05$.

We tested three types of continuum extrapolation; a constant fit with
three or two data points, or a linear extrapolation
\footnote{${\cal O}(g_0^2a/L)$ error is expected from boundary terms in
temporal direction in the SF scheme, which may propagate to
$\alpha_s(M_Z)$ through $\ovl{g}^2(L_{\rm max})$.}.
These results agree with each other and we adopt the constant fit with
three data points for our final results since there is almost no scaling
violation.
Our final results are
\begin{eqnarray}
&&
\alpha_s(M_Z)=0.12047(81)(48)(^{+0}_{-173}),
\label{eqn:alpha5}
\\&&
\Lambda_{\ovl{\rm MS}}^{(5)}=239(10)(6)(^{+0}_{-22}) \;{\rm MeV},
\label{eqn:lambda5}
\end{eqnarray}
where the first parenthesis is statistical error and the second is
systematic error of perturbative matching of different flavors,
which is estimated as a difference between results with three- and two-
loop matching relation for \eqn{eqn:nf2nf-1} and may be overestimated.
The last parenthesis is a difference between the constant and a linear
extrapolation and is a systematic error due to finite lattice spacing for
physical inputs.

\reseteqnum
\section{Conclusion}

We have presented a calculation of the running coupling constant for the
$N_f=2+1$ QCD in the mass independent Schr\"odinger functional scheme in
the chiral limit.
We used seven scales to cover low to high energy
regions and three lattice spacings to take the continuum limit at
each scale.

After tuning $\beta$ and $\kappa$ to fix seven scales in the massless
limit we evaluated the step scaling function in the continuum limit.
We notice that deviation \eqn{eqn:one-loop-deviation} from the continuum
SSF is rather large at one loop for our choice of the Iwasaki gauge
action and the tree level improvement for boundary coefficient
$c_t^{P/R}$.
Since the one loop formula could not reproduce the numerical data except
for very high $\beta\ge10$ we adopted ``two loops'' formula extracted
from numerical data with quadratic fit.
With the ``perturbative'' improvement the SSF shows good scaling
behavior and the continuum limit seems to be taken safely
with a constant extrapolation of the finest two lattice spacings.

We notice that ``two loop'' term in the deviation \eqn{eqn:delta} has
been comparable to that at one loop.
There may be a possibility that higher order perturbative correction
contribute in an non-negligible manner, which may introduce an unestimated
systematic error.
However we consider the probability is not so high
since scaling behavior of the ``two loops'' improved SSF is good as in
Fig.\ref{fig:SSF} and the continuum limit was taken safely.
But a further test may be preferable with a different setup with better
perturbative behavior for the SSF.

With the non-perturbative renormalization group flow we are able to
estimate the renormalization group invariant scale $\Lambda_{\rm QCD}$
and $\alpha_s(M_Z)$ with some physical inputs for energy scale.
The physical scale is introduced from the spectrum simulations of
CP-PACS/JLQCD collaboration \cite{Ishikawa:2007nn} through the hadron
masses $m_\pi$, $m_K$, $m_\Omega$.
From these inputs we evaluated
\eqn{eqn:alpha5} and \eqn{eqn:lambda5},
where all the statistical and systematic errors are included.
Our result is consistent with recent lattice results
\cite{Davies:2008sw,Davies:2008nq,Maltman:2008vj,Allison:2008xk}
and the Particle Data Group average $\alpha_s(M_Z)=0.1176(20)$
\cite{PDG2008} with the systematic error included.

For a future plan a new result is going to be available by the PACS-CS
Collaboration \cite{Aoki:2008sm,Ukita:2008mq} aiming at simulations at
the physical light quark masses down to $m_\pi/m_\rho\approx 0.2$.
This may reveal a systematic error from the physical scale input due to
chiral extrapolation toward light quark masses.
With progress in the physical point simulation expected in the near
future, we are hopeful that a full control of errors in the lattice QCD
determination of the strong coupling constant is in sight.

%
\section*{Acknowledgments}
%
This work is supported in part by Grants-in-Aid of the Ministry
of Education, Culture, Sports, Science and Technology-Japan
 (Nos. 18740130, 18104005, 20340047, 20105001,20105003, 21340049).

\clearpage
\begin{table}[htb]
\begin{center}
\begin{tabular}{|c|c|c|c|c|c|c|c|}
\hline
$\beta$ & $\kappa$ & $L/a$ & $\ovl{g}^2(L)$ & $m_{\rm PCAC}$ &
$2L/a$ & $\ovl{g}^2(2L)$ & $m_{\rm PCAC}$ \\
\hline
$2.15747$ & $0.134249$ & $4$ & $3.4102(99)$ & $-0.00040(21)$ &
$8$ & $5.398(50)$ & $0.040145(29)$\\
$2.34652$ & $0.134439$ & $6$ & $3.415(16)$ & $-0.000003(65)$ &
$12$ & $5.079(89)$ & $0.002629(23)$\\
$2.5$ & $0.133896$ & $8$ & $3.418(19)$ & $0.000020(33)$ &
$16$ & $5.100(143)$ & $0.000462(18)$\\
\hline
$2.5352$ & $0.132914$ & $4$ & $2.6299(29)$ & $0.000112(86)$ &
$8$ & $3.365(26)$ & $0.028468(42)$\\
$2.73466$ & $0.133083$ & $6$ & $2.6292(77)$ & $-0.000015(45)$ &
$12$ & $3.341(49)$ & $0.001633(29)$\\
$2.9$ & $0.132658$ & $8$ & $2.6317(125)$ & $0.000167(27)$ &
$16$ & $3.362(55)$ & $0.000450(13)$\\
\hline
$2.9605$ & $0.131831$ & $4$ & $2.1279(23)$ & $-0.000021(79)$ &
$8$ & $2.553(16)$ & $0.022388(40)$\\
$3.16842$ & $0.131997$ & $6$ & $2.1249(56)$ & $-0.000271(41)$ &
$12$ & $2.5452(257)$ & $0.000960(21)$\\
$3.3$ & $0.131743$ & $8$ & $2.1289(92)$ & $0.000058(27)$ &
$16$ & $2.601(37)$ & $0.000239(12)$\\
\hline
$3.33886$ & $0.131092$ & $4$ & $1.8426(19)$ & $0.000035(78)$ &
$8$ & $2.1191(68)$ & $0.018977(19)$\\
$3.55351$ & $0.131244$ & $6$ & $1.8375(32)$ & $0.000029(28)$ &
$12$ & $2.106(19)$ & $0.000934(17)$\\
$3.7$ & $0.131021$ & $8$ & $1.8403(59)$ & $0.000086(19)$ &
$16$ & $2.165(38)$ & $0.000199(14)$\\
\hline
$3.93653$ & $0.130195$ & $4$ & $1.5248(10)$ & $0.000148(50)$ &
$8$ & $1.7082(54)$ & $0.015474(19)$\\
$4.15042$ & $0.130356$ & $6$ & $1.5300(40)$ & $0.000032(37)$ &
$12$ & $1.692(11)$ & $0.000529(12)$\\
$4.3$ & $0.1302$ & $8$ & $1.5242(35)$ & $-0.000383(15)$ &
$16$ & $1.6959(147)$ & $-0.000295(8)$\\
\hline
$4.74$ & $0.12934$ & $4$ & $1.24874(84)$ & $-0.000098(45)$ &
$8$ & $1.3640(50)$ & $0.012179(21)$\\
$4.94755$ & $0.129495$ & $6$ & $1.2483(15)$ & $0.000166(18)$ &
$12$ & $1.3614(66)$ & $0.0005732(94)$\\
$5.1$ & $0.129376$ & $8$ & $1.2488(28)$ & $0.000028(14)$ &
$16$ & $1.3642(85)$ & $0.000055(6)$\\
\hline
$5.87312$ & $0.128517$ & $4$ & $0.99982(61)$ & $-0.000020(43)$ &
$8$ & $1.0695(30)$ & $0.009380(15)$\\
$6.06879$ & $0.12867$ & $6$ & $1.00130(97)$ & $0.000022(15)$ &
$12$ & $1.0699(44)$ & $0.0002568(82)$\\
$6.2$ & $0.1286$ & $8$ & $1.0006(16)$ & $0.000009(10)$ &
$16$ & $1.0827(70)$ & $-0.000025(6)$\\
\hline
\end{tabular}
\caption{The value of $\beta$ and $\kappa$ to reproduce the same
 physical box size $L$ and near zero PCAC mass.
 The renormalized coupling and PCAC mass at scale $L$ and $2L$ is
 also listed.
}
\label{tab:beta-kappa}
\end{center}
\end{table}

\begin{table}[htb]
\begin{center}
\begin{tabular}{|c|c|c|c|c|c|}
\hline
$\beta$ & $\kappa$ & $L/a$ & \# of confs. & $2L/a$ & \# of confs. \\
\hline
$2.15747$ & $0.134249$ & $4$ & $100000$ & $8$ & $380000$ \\
$2.34652$ & $0.134439$ & $6$ & $120000$ & $12$ & $119200$ \\
$2.5$ & $0.133896$ & $8$ & $134200$ & $16$ & $54400$ \\
\hline
$2.5352$ & $0.132914$ & $4$ & $320000$ & $8$ & $74000$ \\
$2.73466$ & $0.133083$ & $6$ & $144000$ & $12$ & $34100$ \\
$2.9$ & $0.132658$ & $8$ & $122200$ & $16$ & $51200$ \\
\hline
$2.9605$ & $0.131831$ & $4$ & $210000$ & $8$ & $50000$ \\
$3.16842$ & $0.131997$ & $6$ & $110000$ & $12$ & $40400$ \\
$3.3$ & $0.131743$ & $8$ & $74000$ & $16$ & $39600$ \\
\hline
$3.33886$ & $0.131092$ & $4$ & $170000$ & $8$ & $134000$ \\
$3.55351$ & $0.131244$ & $6$ & $170000$ & $12$ & $35300$ \\
$3.7$ & $0.131021$ & $8$ & $98000$ & $16$ & $23200$ \\
\hline
$3.93653$ & $0.130195$ & $4$ & $230000$ & $8$ & $86000$ \\
$4.15042$ & $0.130356$ & $6$ & $170000$ & $12$ & $47800$ \\
$4.3$ & $0.1302$ & $8$ & $122000$ & $16$ & $41600$ \\
\hline
$4.74$ & $0.12934$ & $4$ & $170000$ & $8$ & $40000$ \\
$4.94755$ & $0.129495$ & $6$ & $150000$ & $12$ & $51200$ \\
$5.1$ & $0.129376$ & $8$ & $86000$ & $16$ & $52000$ \\
\hline
$5.87312$ & $0.128517$ & $4$ & $110000$ & $8$ & $40000$ \\
$6.06879$ & $0.12867$ & $6$ & $153000$ & $12$ & $41600$ \\
$6.2$ & $0.1286$ & $8$ & $98000$ & $16$ & $30800$ \\
\hline
\end{tabular}
\caption{Number of configurations for each run.}
\label{tab:beta-kappa-conf}
\end{center}
\end{table}

\begin{table}[htb]
\begin{center}
\begin{tabular}{|c|c|c|}
\hline
$L/a$ & $\delta_{1G}$ & $\delta_{1Q}$\\
\hline
$4$ & $-0.02096$ & $-0.00470$\\
$6$ & $-0.01922$ & $-0.00329$\\
$8$ & $-0.01499$ & $-0.00248$\\
$10$ & $-0.01241$ & $-0.00197$\\
$12$ & $-0.01064$ & $-0.00163$\\
\hline
\end{tabular}
\caption{Perturbative improvement factor at one loop level for tree level
 improved $c_t^{P/R}$.
Pure gauge contribution $\delta_{1G}$ is taken from
 Ref.~\cite{Takeda:2003he} and quark contribution $\delta_{1Q}$ from
\cite{Takeda2009}.}
\label{tab:delta}
\end{center}
\end{table}

\begin{table}[htb]
\begin{center}
\begin{tabular}{|c|c|c|c|c|c|c|c|}
\hline
$\beta$ & $\kappa$ & $L/a$ & $\ovl{g}^2(L)$ & $m_{\rm PCAC}$ &
$2L/a$ & $\ovl{g}^2(2L)$ & $m_{\rm PCAC}$ \\
\hline
$10$ & $0.1270893$ & $4$ & $0.58565(34)$ & $-0.000093(44)$ &
$8$ & $0.6055(16)$ & $0.004695(20)$\\
$20$ & $0.1260654$ & $4$ & $0.29543(17)$ & $-0.000059(32)$ &
$8$ & $0.29943(57)$ & $0.001720(18)$\\
$40$ & $0.1255571$ & $4$ & $0.14876(23)$ & $-0.000051(37)$ &
$8$ & $0.14896(60)$ & $0.000182(25)$\\
$60$ & $0.1253871$ & $4$ & $0.099336(40)$ & $-0.000051(33)$ &
$8$ & $0.09986(25)$ & $-0.000293(22)$\\
$80$ & $0.1253023$ & $4$ & $0.074654(11)$ & $-0.000043(13)$ &
$8$ & $0.07485(12)$ & $-0.000567(12)$\\
$100$ & $0.1252498$ & $4$ & $0.059801(24)$ & $0.000030(15)$ &
$8$ & $0.059927(85)$ & $-0.000636(7)$\\
\hline
$10$ & $0.1272305$ & $6$ & $0.59707(44)$ & $0.000012(14)$ &
$12$ & $0.6229(22)$ & $-0.000090(9)$\\
$20$ & $0.1261216$ & $6$ & $0.29775(33)$ & $0.000021(19)$ &
$12$ & $0.30322(92)$ & $-0.000310(10)$\\
$40$ & $0.1255700$ & $6$ & $0.149219(45)$ & $0.000002(4)$ &
$12$ & $0.15031(33)$ & $-0.000433(6)$\\
$60$ & $0.1253863$ & $6$ & $0.099664(47)$ & $0.000019(9)$ &
$12$ & $0.10012(19)$ & $-0.000460(7)$\\
$80$ & $0.1252948$ & $6$ & $0.074730(99)$ & $0.000005(14)$ &
$12$ & $0.07532(37)$ & $-0.000499(13)$\\
$100$ & $0.1252397$ & $6$ & $0.059844(44)$ & $0.000000(5)$ &
$12$ & $0.06029(18)$ & $-0.000493(6)$\\
\hline
$10$ & $0.1272310$ & $8$ & $0.6051(12)$ & $-0.000039(17)$ &
$16$ & $0.6296(35)$ & $-0.000146(6)$\\
$20$ & $0.1261176$ & $8$ & $0.29971(49)$ & $-0.000024(14)$ &
$16$ & $0.3054(22)$ & $-0.000233(11)$\\
$40$ & $0.1255626$ & $8$ & $0.14948(27)$ & $-0.00000(1)$ &
$16$ & $0.15158(65)$ & $-0.000261(79)$\\
\hline
\end{tabular}
 \caption{The renormalized coupling and PCAC mass at scale $L$ and $2L$
 to derive the SSF at high $\beta\ge10$.}
\label{tab:pt-beta-kappa}
\end{center}
\end{table}

\begin{table}[htb]
\begin{center}
\begin{tabular}{|c||c|c|}
\hline
$L/a$ & $d_1=\delta_1$ & $d_2$ \cr
\hline
$4$ & $-0.03506$ & $0.013690$ \cr 
$6$ & $-0.02909$ & $0.008307$ \cr 
$8$ & $-0.02243$ & $0.004936$ \cr 
\hline
\end{tabular}
 \caption{Coefficients of the quadratic fit of the deviation
 $\delta(u,a/L)$.}
\label{tab:polynomial-fit}
\end{center}
\end{table}

\begin{table}[htb]
\begin{center}
\begin{tabular}{|c|c|c|c|c|}
\hline
$u$ & $\sigma(u)$ & $\Sigma^{\rm (2)}(u,1/8)$ & $\Sigma^{\rm (2)}(u,1/6)$
 & $\Sigma^{\rm (2)}(u,1/4)$\cr
\hline
$1.0006$ & $1.0947(39)$ & $1.1020(74)$ & $1.0918(46)$ & $1.0939(32)$ \cr
$1.2488$ & $1.3937(56)$ & $1.3924(93)$ & $1.3945(71)$ & $1.3954(53)$ \cr
$1.5242$ & $1.7380(93)$ & $1.736(16)$ & $1.739(11)$ & $1.7450(57)$ \cr
$1.8403$ & $2.175(18)$ & $2.220(41)$ & $2.165(20)$ & $2.1549(75)$ \cr
$2.1289$ & $2.632(23)$ & $2.669(41)$ & $2.615(28)$ & $2.587(17)$ \cr
$2.6317$ & $3.426(39)$ & $3.447(61)$ & $3.411(52)$ & $3.359(27)$ \cr
$3.4178$ & $5.127(80)$ & $5.20(15)$ & $5.098(95)$ & $5.206(52)$ \cr
\hline
\end{tabular}
\caption{The $O(a)$ improved SSF $\Sigma^{\rm (2)}(u,a/L)$ at ``two
 loop'' level for three lattice spacings $a/L=1/4$, $1/6$, $1/8$.
 Corrections are made such that the PACS mass $m=0$ and the renormalized
 coupling $u$ at smaller box to be the same value for each lattice
 spacings.
 The SSF $\sigma(u)$ in the continuum is also listed, which is given by
 a constant fit of two data at finest lattice spacings $1/6$, $1/8$.}
\label{tab:ssf}
\end{center}
\end{table}

\begin{table}[htb]
\begin{center}
\begin{tabular}{|c|c|c|c|}
\hline
$\beta$ & $1.83$ & $1.90$ & $2.05$ \\
\hline
$a$ (fm) & $0.1209(16)$ & $0.0982(19)$ & $0.0685(26)$ \\
\hline
\end{tabular}
\caption{Lattice spacing $a$ from large scale simulation
 \cite{Ishikawa:2007nn}.}
\label{tab:lattice-spacing}
\end{center}
\end{table}

\begin{table}[htb]
\begin{center}
\begin{tabular}{|c|c|c|c|c|}
\hline
$\beta$ & $\kappa$ & $L_{\rm max}/a$ & $\ovl{g}^2(L_{\rm max})$
 & $m_{\rm PCAC}$ \\
\hline
$1.83$ & $0.13608455$ & $4$ & $5.565(54)$ & $0.00015(56)$ \\
$1.83$ & $0.138685$ & $6$ & $7.79(20)$ & $-0.02181(56)$ \\
\hline
$1.90$ & $0.1355968$ & $4$ & $4.695(23)$ & $-0.00039(28)$ \\
$1.90$ & $0.1372766$ & $6$ & $6.71(16)$ & $0.00099(38)$ \\
$1.90$ & $0.137659$ & $8$ & $9.15(60)$ & $-0.00547(52)$ \\
\hline
$2.05$ & $0.1347342$ & $4$ & $3.806(13)$ & $-0.00023(22)$ \\
$2.05$ & $0.1359925$ & $6$ & $4.740(79)$ & $0.00022(23)$ \\
$2.05$ & $0.136115987$ & $8$ & $6.01(21)$ & $0.00026(15)$ \\
\hline
\end{tabular}
 \caption{The renormalized coupling and PCAC mass at $\beta=1.83$,
 $1.90$, $2.05$ used in the large scale simulation.
 The values of $\kappa$'s are tuned to reproduce $m_{\rm PCAC}=0$ except
 for the case that the coupling apparently exceeds $5.13$.}
\label{tab:lmax}
\end{center}
\end{table}

\begin{table}[htb]
\begin{center}
\begin{tabular}{|c|c|c|c|c|c|c|}
\hline
$\beta$ & $L_{\rm max}/a$ & $1/L_{\rm max}$ (MeV)
 & $\Lambda_{\rm SF}^{(3)}L_{\rm max}$ & $\Lambda_{\rm SF}^{(3)}$ (MeV)
 & $\Lambda_{\ovl{\rm MS}}^{(3)}$ (MeV) \cr
\hline
$1.83$ & $4$ & $408.0(5.4)$ & $0.355(18)$ & $144.8(7.8)$ & $378(20)$ \cr
$1.90$ & $4$ & $502.3(9.7)$ & $0.286(16)$ & $143.6(8.5)$ & $375(22)$ \cr
$2.05$ & $4$ & $720(27)$ & $0.202(12)$ & $145(10)$ & $379(26)$ \cr
$2.05$ & $6$ & $480(18)$ & $0.290(16)$ & $139.2(9.4)$ & $364(24)$ \cr
$2.05$ & $8$ & $360(14)$ & $0.385(20)$ & $138.7(8.9)$ & $362(23)$ \cr
\hline
\end{tabular}
\caption{The RGI scale $\Lambda_{\rm SF}^{(3)}$ for three flavors in
 the SF scheme and $\Lambda_{\ovl{\rm MS}}^{(3)}$ in the $\ovl{\rm MS}$
 scheme.}
\label{tab:lambda}
\end{center}
\end{table}

\begin{table}[htb]
\begin{center}
\begin{tabular}{|c|c|c|c|c|c|}
\hline
$\beta$ & $L_{\rm max}/a$ & $\alpha_s(M_Z)$
 & $\Lambda_{\ovl{\rm MS}}^{(5)}$ (MeV) & $\alpha_s(M_Z)$
 & $\Lambda_{\ovl{\rm MS}}^{(5)}$ (MeV)\cr
\hline
$1.83$ & $4$ & $0.1208(13)$ & $243(17)$ & $0.1203(12)$ & $237(16)$ \cr
$1.90$ & $4$ & $0.1206(14)$ & $240(18)$ & $0.1201(14)$ & $234(17)$ \cr
$2.05$ & $4$ & $0.1208(17)$ & $244(22)$ & $0.1204(16)$ & $237(21)$ \cr
$2.05$ & $6$ & $0.1198(16)$ & $231(20)$ & $0.1194(15)$ & $225(19)$ \cr
$2.05$ & $8$ & $0.1198(15)$ & $230(19)$ & $0.1193(14)$ & $224(18)$ \cr
\hline
\end{tabular}
\caption{The strong coupling $\alpha_s(M_Z)$ at $\mu=M_Z$ and the RGI
 scale $\Lambda_{\ovl{\rm MS}}^{(5)}$ for five flavors.
 Those in the third and the fourth column are derived with three loops
 formula for \eqn{eqn:nf2nf-1}.
 Those in the fifth and the sixth are from two loops formula.}
\label{tab:alphas}
\end{center}
\end{table}

\clearpage

\end{document}